\documentclass[preprint, prb, showpacs, superscriptaddress]{revtex4}
\usepackage{epsfig}
\begin{document}
\topmargin-1.0cm

\title {
Electronic structures and optical properties of layered
perovskites Sr$_2$$M$O$_4$ ($M$=Ti, V, Cr, and Mn): An $ab$ $ $
$initio$ study }

\author{Hongming Weng}\email[Corresponding author Electronic address:]{hongming@imr.edu}
\affiliation {Institute for Materials Research, Tohoku University,
Sendai 980-8577, Japan}
\author{Y. Kawazoe}
\affiliation {Institute for Materials Research, Tohoku University,
Sendai 980-8577, Japan}
\author {Xiangang Wan}
\affiliation {Group of Computational Condensed Matter Physics,
National Laboratory of Solid State Microstructures and Dept. of
Physics, Nanjing University, Nanjing 210093, P. R. China}
\author {Jinming Dong}
\affiliation {Group of Computational Condensed Matter Physics,
National Laboratory of Solid State Microstructures and Dept. of
Physics, Nanjing University, Nanjing 210093, P. R. China}
\date{\today}

\begin{abstract}

A series of layered perovskites Sr$_2$$M$O$_4$ ($M$=Ti, V, Cr, and
Mn) is studied by $ab $ $initio$ calculations within generalized
gradient approximation (GGA) and GGA+$U$ schemes. The total
energies in different magnetic configurations, including the
nonmagnetic, ferromagnetic, the layered antiferromagnetic with
alternating ferromagnetic plane and the staggered in-plane
antiferromagnetic (AFM-II) order, are calculated. It is found that
Sr$_2$TiO$_4$ is always a nonmagnetic band insulator. For
Sr$_2$MnO$_4$, both GGA and GGA+$U$ calculations show that the
insulating AFM-II state has the lowest total energy among all the
considered configurations. For $M$=V and Cr, the GGA is not enough
to give out the insulating AFM-II states and including the on-site
electron-electron correlation effect $U$ is necessary and
efficient. The AFM-II state will have the lowest total energy in
both cases when $U$ is larger than a critical value. Further, the
optical conductivity spectra are calculated and compared with the
experimental measurements to show how well the ground state is
described within the GGA or GGA+$U$. The results indicate that $U$
is overestimated in Sr$_2$VO$_4$ and Sr$_2$CrO$_4$. To make up
such a deficiency of GGA+$U$, the contributions from proper
changes in the ligand field, acting cooperatively with $U$, are
discussed and shown to be efficient in Sr$_2$CrO$_4$.
\end{abstract}


\maketitle

\section{introduction} \label{introduction}
Perovskite transition-metal oxides have been intensively
studied\cite{imada} due to their various intriguing physical
properties, such as the ferroelectricity, complex charge, spin
and/or orbital order, and some combined properties including
magnetoelectric multiferroics\cite{spaldin} and colossal
magnetoresistance.\cite{tokura} In this family, the
two-dimensional layered perovskite in K$_2$NiF$_4$ structure is
particularly interesting because of the high-$T_c$
superconductivity in cuprates based on La$_2$CuO$_4$ and the novel
charge, spin, and/or orbital stripes in nickelates and
manganites.\cite{phystoday} Compared with the quite early success
in the synthesis of ABO$_3$ like perovskites, only recently has a
series of layered perovskites Sr$_2$$M$O$_4$ with transition-metal
$M$=Ti, V, Cr, Mn, and Co been successfully synthesized in
single-crystalline thin films.\cite{srmo} Thus, the variation of
their electronic structures could be systematically studied as the
occupation of $d$ orbitals increases from empty ($d^0$ for
Ti$^{4+}$) to half-filled ($d^5$ for Co$^{4+}$).\cite{srmo} In the
$d^0$ case, Sr$_2$TiO$_4$ is always a nonmagnetic insulator. For
$M$=V, the surrounding elongated octahedral crystal field is
thought to make the single $d^1$ electron stay in the doubly
degenerate $d_{xz+yz}$ orbitals and thus result in metallic
features, but experimentally Sr$_2$VO$_4$ is found to be an
antiferromagnetic (AFM) insulator.\cite{srmo} Since the singly
occupied degenerate $d_{xz+yz}$ orbitals are active in the orbital
degree of freedom, a kind of spin and/or orbital
ordering\cite{srmo, path} has been proposed to explain its
insulating AFM ground state. In the $M$=Cr case, two $d$ electrons
are supposed to occupy the doubly degenerate $d_{xz+yz}$ orbitals
and are expected to be a simple Mott-Hubbard insulator in stark
contrast to the metallic cubic perovskite SrCrO$_3$.\cite{srcro3}
Ignoring the empty $e_g$ orbitals, Sr$_2$CrO$_4$ has a dual
relation to Sr$_2$RuO$_4$, which has two holes instead of two
electrons in the narrow $t_{2g}$ orbitals and possesses
superconductivity.\cite{srruo} Three $d$ electrons in
Sr$_2$MnO$_4$ will fully occupy the lower Hubbard bands of the
$t_{2g}$ orbitals, making the system an AFM insulator. All the
systems above only have $t_{2g}$ electrons in the states around
the Fermi level, but for $M$=Co it is totally different from the
others by having electrons in the $e_g$ orbitals and possessing
metallic FM orders.\cite{srmo, srcoo} All these systematical
variations of electronic structures are revealed by optical
conductivity measurements.\cite{srmo} The anisotropy of the
polarized optical conductivity due to the two-dimensional $M$O$_2$
sheets in the layered perovskite structures is also clearly shown
by measurements with $E$$\perp$$c$ and $E$$\parallel$$c$,
respectively.

Though it is now found quite difficult to predict the correct spin
and orbital state in the $t_{2g}$ system and more and more
researchers agree that methods beyond the local density
approximation (LDA) seem to be necessary, the conventional
description within the LDA and LDA+$U$ is still important and key
as basic knowledge prior to the other methods.\cite{solovyev}
Sr$_2$VO$_4$ has been already carefully investigated by
first-principles calculations\cite{pickett, singh} within local
spin density approximation (LSDA). But the LSDA calculation failed
in reproducing its AFM insulator state, and a +$U$ calculation has
not been tried. To the best of our knowledge, no first-principles
calculations have been performed on Sr$_2$CrO$_4$ so far. Wang
{\it et al.}\cite{wang} studied Sr$_2$MnO$_4$ just within the LSDA
and found it is an AFM insulator. For Sr$_2$CoO$_4$, both LSDA and
LSDA+$U$ calculations\cite{srcoo, lee} can successfully describe
this system and show that the strong hybridization between the Co
$3d$ and O $2p$ orbitals makes this $d^5$ system a ferromagnetic
(FM) metal. Therefore, in this paper, we put emphasis on the so
called $t_{2g}$ system with $M$=V, Cr, Mn and will not cover
Sr$_2$CoO$_4$ again. We use the generalized gradient approximation
(GGA) instead of the LSDA in order to improve the accuracy for
spin-polarized calculation.\cite{gga} The on-site
electron-electron correlation effect $U$ is also taken into
account for these systems. To make a complete theoretical research
and comparison, various magnetic configurations for each case are
studied. Though the three-dimensional ferromagnetism and layered
AFM state with alternating ferromagnetic $M$O$_2$ sheets (AFM-I)
have been ruled out by experimental measurements on thin-film
samples,\cite{srmo} both of them are calculated here. In layered
perovskite structure, the typical and simplest magnetic
configuration, staggered in-plane AFM order (AFM-II),\cite{weng}
is also included, which is the most important configuration for
understanding the underlying physics in these systems despite the
fact that some other complex spin and/or orbital orderings may
exist in the $M$=V or Cr case.\cite{srmo, path} $U$ is found to
play an important role in both Sr$_2$VO$_4$ and Sr$_2$CrO$_4$
where the GGA is insufficient for predicting the AFM insulating
ground states. The effects of the changes in ligand field, which
might influence the $U$ value, are also discussed. Finally, the
optical conductivity spectra are calculated and compared with the
experimental measurements to show the efficiency of the GGA and
GGA+$U$ schemes in describing these systems.

\section{Methodology} \label{Methodology}
The Vienna ab initio simulation package (VASP) plane-wave
code\cite{vasp} has been employed in our calculations. The
projector augmented-wave (PAW) version of
pseudopotential\cite{paw} is used to describe the electron-ion
interaction. The exchange correlation term is treated within the
GGA as parametrized by Perdew {\it et al}.\cite{gga} The approach
of Dudarev {\it et al.} for adding $U$ is used in the GGA+$U$
calculation, in which only one effective on-site Coulomb
interaction parameter is needed.\cite{dudarev} For all
Sr$_2$$M$O$_4$, the lattice constants are taken from experimental
measurements of thin-film samples\cite{srmo} as shown in Table I.
There is an orthorhombic deviation from tetragonal symmetry in
each case. The largest distortion is in the $M$=Cr case with
$a/b$=1.023. In the other cases, $a/b$ are all nearly 1.0. In the
first approximation,\cite{srcoo} such an orthorhombic deviation
from tetragonal symmetry is ignored and all are assumed to have
typical K$_2$NiF$_4$ structure in our calculation. The internal
atomic coordinates are fully optimized until the force on the atom
is less than 0.002 eV/\AA.$ $ For simulating the AFM-II state, a
cell as large as $\sqrt{2}\times\sqrt{2}\times1$ of the
conventional K$_2$NiF$_4$ unit cell is needed, which is also used
for other states in order to accurately compare their total
energies.

The interband optical responses are calculated within the
electric-dipole approximation using the Kubo formula\cite{kubo}
implemented by Furthm\"{u}ller,\cite{fur} in which the imaginary
part of the dielectric function can be expressed as
\[\begin{array}{l}
\varepsilon_2(\omega)=\frac{8\pi^2e^2}{\omega^2m^2V}
\sum\limits_{c,v} {\sum\limits_k {\vert < c,k\vert {\rm {\bf
\hat{e}}} \cdot } } {\rm {\bf p}}\vert v,k> \vert^2 \\
\times
\delta [E_c (k) - E_v (k) - \hbar \omega ], \\
\end{array}\]
\noindent where $c$ and $v$ represent the conduction and valence
bands, respectively. $|c,k>$ ($|v,k>$) and $E_c$ ($E_v$) are the
eigenstate and eigenvalue of the conduction (valence) band
obtained from VASP calculations, respectively. {\bf p} is the
momentum operator, ${\rm {\bf \hat {e}}}$ is the electric field
vector of the incident photon, denoting polarization of the light,
and $\omega$ is its frequency. An $8\times8\times8$ grid is used
to sample the Brillouin zone for integration over $k$ space with
the linear tetrahedron scheme improved by Bl\"{o}chl {\it et
al.}~\cite{blochl} The energy cutoff of the plane wave is 520 eV.
The convergence on $k$ points and energy cutoff is carefully
checked.

\section{results and discussions} \label{result}

\subsection{Sr$_2$TiO$_4$}\label{Sr2TiO4}
The lattice constants of the Sr$_2$$M$O$_4$ are taken from the
experimental measurements and only the internal positions of the
apical oxygen (O$_a$) and the Sr atoms are needed to be relaxed
due to the symmetry of the crystal structure. The obtained results
are listed in Table I for Sr$_2$TiO$_4$ together with $M$=V, Cr,
and Mn. For Sr$_2$TiO$_4$, the internal positions of O$_a$ and Sr
are 0.159 50 and 0.145 89, being comparable with those obtained by
other LDA calculations,\cite{srtio} 0.160 and 0.145, respectively.
In all cases, the oxygen octahedron surrounding the $M$ ion has a
similar Jahn-Teller (JT) distortion with an elongation along the
$c$ axis. The distortion degree is defined as $J$=$d_c$/$d_{ab}$
with $d_c$ ($d_{ab}$) being the $M$-O bond length along the $c$
axis (in the $ab$ plane), as listed in Table I. Among them, $M$=Cr
and Mn have the largest distortion and $M$=V has the smallest.

\begin{table}
\caption{The lattice constants taken from the experimental
measurements and the theoretically optimized internal coordinates
relative to the $c$ lattice for the apical oxygen (O$_a$) and Sr
atoms. $J$=$d_{c}/d_{ab}$ is the Jahn-Teller distortion degree
with $d_{c}$ ($d_{ab}$) being the $M$-O bond length along the $c$
axis (in the $ab$ plane). The length unit is in angstroms.}
\label{tab1}
\begin{tabular}{lcccc}
\hline \hline
            & Ti & V  & Cr & Mn \\
\hline
a &  3.866  & 3.832 &  3.756 &  3.755 \\
c &  12.60  & 12.59 &  12.59 &  12.62 \\
O$_a$ & 0.15950 & 0.15781 & 0.15899 & 0.15858 \\
Sr & 0.14589 & 0.14486 & 0.14673 & 0.14646 \\
$d_{ab}$ & 1.93 & 1.92    & 1.88    & 1.88   \\
$d_c$ & 2.01 & 1.99 & 2.00    & 2.00     \\
$J$   & 1.041 & 1.036 &  1.063 & 1.063   \\
\hline \hline
\end{tabular}
\end{table}

\begin{figure}
\centering
\includegraphics[width=0.8\textwidth]{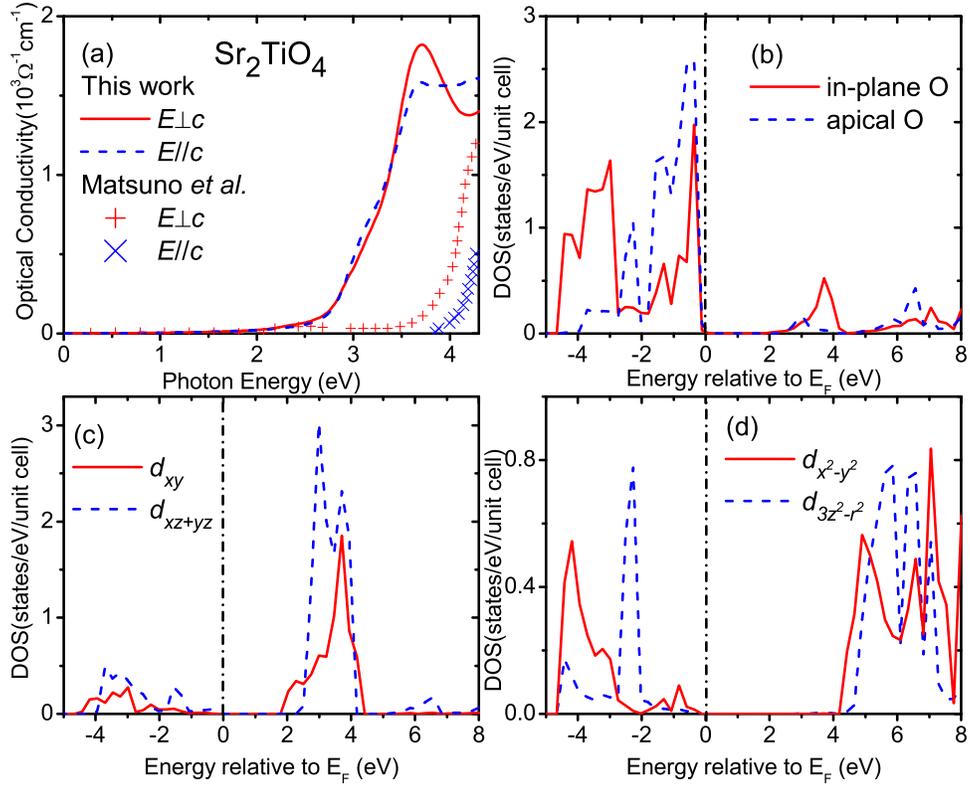}
\caption{(Color online) Calculated results within the GGA for
Sr$_2$TiO$_4$. (a) Optical conductivity spectra (lines), (b)
projected partial DOS for in-plane (solid) and apical (dashed)
oxygen $2p$ bands, and (c) $t_{2g}$ and (d) $e_{g}$ bands of the
Ti $3d$ electrons. In (a), the experimental spectra (symbols, Ref.
\onlinecite{srmo}) are plotted in symbols. The vertical
dash-dotted line indicates the Fermi energy level.}\label{fig1}
\end{figure}

The Sr$_2$TiO$_4$ system is quite simple. It is a band insulator
as shown by the density of states (DOS) in Fig. 1. The valance and
conduction bands are mostly composed of the oxygen $2p$ and Ti
$3d$ bands, respectively. The width of the $t_{2g}$ bands is much
narrower than that of the $e_g$ bands because of their different
bonding type with the oxygen $p$ bands. The in-plane $d_{xy}$ and
$d_{x^2-y^2}$ bands are wider than the out-of-plane $d_{xz+yz}$
and $d_{3z^2-r^2}$ bands, respectively, due to the elongated
oxygen octahedra, which also make the in-plane oxygen $p$ bands
wider than that of the apical one. The calculated polarized
optical conductivity spectra with electric field $E$ parallel
($E$$\parallel$$c$) and perpendicular ($E$$\perp$$c$) to the $c$
axis are shown in Fig. 1(a). The experimental spectra\cite{srmo}
are also plotted for easy comparison. Obviously, the calculated
optical gap is underestimated due to the well-known shortcomings
of the GGA.\cite{ggashort} The anisotropy of the spectra
originating from the two-dimensional layered structure is also
observed. The shoulder near 3.0 eV in the calculated
$E$$\parallel$$c$ spectrum is contributed by transitions from the
occupied apical oxygen $2p$ bands to the empty $d_{xz+yz}$
orbitals centered around 3.0 eV above the Fermi level. Such a kind
of transition is not observed when $E$$\perp$$c$. In both the
$E$$\perp$$c$ and $E$$\parallel$$c$ cases, the first high peak
around 3.7 eV is attributed to the transition from in-plane and
apical oxygen $2p$ orbitals to the Ti $d_{xy}$ and $d_{xz+yz}$
orbitals, respectively. When $E$$\parallel$$c$, this peak is quite
broader since, just below the Fermi level, the apical oxygen $p$
DOS structure is wider than that of the in-plane oxygen, as shown
in Fig. 1(b).

\subsection{Sr$_2$VO$_4$}\label{SrVO4}
The experimental parameters for O$_a$ and Sr in Sr$_2$VO$_4$ are
0.157 78 and 0.145 62,\cite{jsol} respectively, comparable with
our GGA results, 0.157 81 and 0.144 86. Sr$_2$VO$_4$ has the
smallest JT distortion among all the systems. In order to search
for the possible ground state for $M$=V, Cr, and Mn, we have
calculated the total energies of the optimized structure in
various magnetic configurations. The obtained total energy
relative to the FM state by GGA calculations is shown in Table II.
For $M$=V, the AFM-I state has the lowest energy, which is
inconsistent with the experimental analysis since the FM order in
the $ab$ plane would give rise to the metallic optical
conductivity when $E$$\perp$$c$. In the left panel of Fig. 2, the
total DOS of Sr$_2$VO$_4$ in the AFM-I state clearly shows
metallic features. In the picture of JT-distortion, the doubly
degenerate $d_{xz+yz}$ orbitals are lower than the $d_{xy}$
orbital, but all of them have contributions to the DOS at the
Fermi level although there is only one valence $d$ electron. This
is very similar to that in the $d^4$ system of
Sr$_2$RuO$_4$.\cite{anisimov} It should be mentioned that in
Sr$_2$VO$_4$, though the AFM-II configuration is initialized for
calculations, the electronic charge density always converges to
the NM state and the AFM-II state can not be achieved in our GGA
calculation. In Table II, the AFM-II energy for $M$=V is in fact
that of the NM state. Pickett {\it et al.} also tried in LSDA
calculations to search for the AFM state of Sr$_2$VO$_4$ but
failed.\cite{pickett}

\begin{table}
\caption{The calculated total energies of Sr$_2$$M$O$_4$ in NM,
AFM-I and AFM-II states within GGA. All the energies are given in
relative to that of FM state and the unit is meV/cell.}
\label{tab2}
\begin{tabular}{lcccc}
\hline \hline
            & V & Cr  & Mn \\
\hline
NM     &  280  & 1252 &  3312  \\
AFM-I  & -6    & -29  & -38    \\
AFM-II & 280   & 297  & -629  \\
\hline \hline
\end{tabular}
\end{table}

\begin{figure}
\centering
\includegraphics[width=0.8\textwidth]{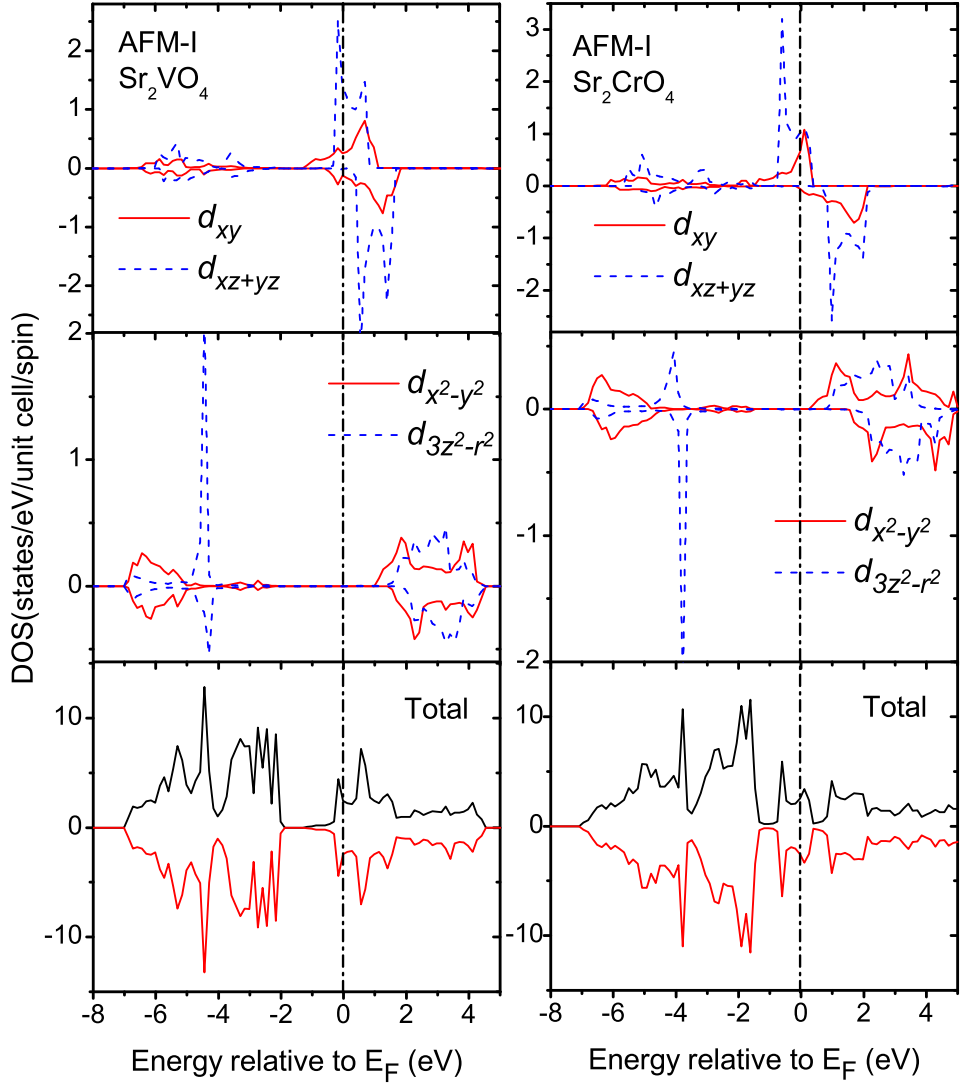}
\caption{(color online) The calculated DOS of the AFM-I state for
$M$=V (left panel) and Cr (right panel) within the GGA. The
vertical dash-dotted line indicates the Fermi energy
level.}\label{fig2}
\end{figure}

For a $d$-electron system, usually the strong electron-electron
correlation $U$ is included in order to remedy the deficiency of
the GGA. In addition, $U$ had been used for predicting the correct
energy order in the $t_{2g}$ system YTiO$_3$, where the GGA
failed.\cite{sawada} In order to determine the proper value of
$U$, a series of calculations on different magnetic states for
Sr$_2$VO$_4$ with different $U$ have been performed. The
variations of total energies in different configurations relative
to the FM state when $U$ is increased are plotted in the left
panel of Fig. 3. The AFM-I state is almost degenerate with the FM
one, indicating that the coupling between the VO$_2$ sheets is
very small. The total energy of the AFM-II state is the lowest
when $U$ is larger than a critical value $U_c$, about 2.08 eV.

The evolution of the $t_{2g}$ DOS with increasing the $U$
parameter in the AFM-II state for $M$=V is plotted in the left
panel of Fig. 4. As $U$ increases, the occupation of the $d$
electron on the V ion site will be slowly shifted from the
$d_{xz+yz}$ to the $d_{xy}$ orbital. Due to the separation of
$d_{xy}$ from $d_{xz+yz}$, a metal-to-insulator transition happens
near $U$=2.0 eV. The $d$ electron eventually occupies the spin-up
$d_{xy}$ orbital instead of the doubly degenerate $d_{xz+yz}$
orbitals though the latter should be lower in energy according to
the picture of elongated JT distortion. This kind of occupancy
will destroy any possible orbital ordering although recently Imai
{\it et al.} proposed a kind of spin and orbital order in this
system.\cite{path} In fact, in the FM state, the doubly degenerate
$d_{xz+yz}$ orbitals occupied by one $d$ electron will cause FM
instability as found by Pickett {\it et al.}\cite{pickett}

Since the N\'eel temperature of Sr$_2$VO$_4$ is reported to be
about 47 K,\cite{srvo} the energy difference between the AFM-II
and FM states should not be too large if using the popular
spin-1/2 two-dimensional Heisenberg model to estimate the N\'eel
temperature.\cite{wan} Therefore, in the calculation of optical
conductivity spectra we roughly take the $U$ value of 2.1 eV,
which is slightly larger than the critical value. The obtained
optical conductivity spectra as well as the experimental ones are
shown in Fig. 5 together with partial DOS for oxygen and V. From
the DOS, it can be seen that the fundamental gap is about 0.1 eV,
but the optical gap, obtained from the calculated optical
conductivity spectra, is as high as about 2.0 eV. Obviously, the
transitions between the occupied $d_{xy}$ orbital and the empty
$d_{xz+yz}$ orbitals, which constitute the fundamental gap, are
forbidden for both $E$$\perp$$c$ and $E$$\parallel$$c$. This is
reasonable and consistent with the optical selection rule of
$d-d^*$ transitions.\cite{njp7147} According to this rule, in the
AFM-II configuration, the transition from the occupied spin-up
$d_{xy}$ orbital to the nearest unoccupied spin-up $d_{xy}$ one is
allowed only for $E$$\perp$$c$, which is the so-called
Mott-Hubbard transition.\cite{srmo} Clearly, such a kind of
transition contributes to the 2.0 eV optical gap and the shoulder
around 2.75 eV in the calculated $E$$\perp$$c$ spectrum. However,
both the optical gap and the transition energy are overestimated
by using such a large $U$ of 2.1 eV as compared with the
experimentally suggested value $\sim$1.0 eV as indicated by the
peak around 1.0 eV in the experimental $E$$\perp$$c$
spectrum.\cite{srmo} The peak around 3.0 eV in the calculated
spectrum for $E$$\perp$$c$ can be attributed to the transition
from the in-plane oxygen $2p$ bands to the empty $d_{xz+yz}$
orbitals. This peak corresponds to the experimental one also near
3.0 eV, which is called the charge transfer gap.\cite{srmo} When
$E$$\parallel$$c$, due to the existence of SrO layers, the
intersite $d-d^*$ Mott-Hubbard transition is negligible and not
seen. The charge transfer gap in this case is due to the
transition from the apical oxygen $2p$ bands to the $d_{xz+yz}$
orbitals. The resulting broad and multipeak structure around 3.0
eV in the calculated spectrum is a little lower than the
corresponding experimental one.

\subsection{Sr$_2$CrO$_4$}\label{SrCrO4}
The optimized internal position parameters for O$_a$ and Sr in
Sr$_2$CrO$_4$ are 0.158 99 and 0.146 73, respectively.
Sr$_2$CrO$_4$ has a larger JT distortion than Sr$_2$VO$_4$. Within
the GGA, it has similar results as Sr$_2$VO$_4$ does. The AFM-I
state has the lowest total energy, and its DOS shown in right
panel of Fig. 2 also possesses metallic features. Similar to
Sr$_2$VO$_4$ and Sr$_2$RuO$_4$, both $d_{xy}$ and doubly
degenerate $d_{xz+yz}$ orbitals have contributions to the DOS at
Fermi level. In Sr$_2$CrO$_4$, the Fermi level is a little higher
than that in Sr$_2$VO$_4$ due to one more $d$ electron.

As done in Sr$_2$VO$_4$, the electron-electron correlation effect
$U$ is taken into account and the total energy dependence on the
$U$ value in various magnetic configurations is plotted in the
right panel of Fig. 3. The energy difference between the AFM-I and
FM state is quite larger than that in Sr$_2$VO$_4$. It seems that
the coupling between CrO$_2$ sheets is quite stronger than that
between VO$_2$ ones. When $U$ is larger than the critical value
$U_c$, about 3.36 eV, the total energy of AFM-II state is the
lowest among all the states. In the right panel of Fig. 4, the
variation of Cr $t_{2g}$ partial DOS with $U$ is also plotted.
Different from that in Sr$_2$VO$_4$, as $U$ increases, the two $d$
electrons on the Cr$^{4+}$ ion will occupy the $d_{xz+yz}$ orbital
more and more, and eventually the $d_{xy}$ orbital is left empty.
The metal-insulator transition will happen at a higher $U$ value
of about 5.0 eV (not shown here), where the $d_{xy}$ orbital is
totally separated from the $d_{xz+yz}$. Here, the $d_{xz+yz}$
orbitals are lower than the $d_{xy}$, which is consistent with the
picture of the elongated JT distortion in the oxygen octahedron.
Therefore, in the $M$=V and Cr cases, the effects of JT distortion
and $U$ are totally different. In the former case, $U$ acts
against the JT distortion, while in the latter case, they work
cooperatively. This can also be seen from the forces on the apical
oxygen atoms. Before the metal-insulator transition, as the $U$
increases, the apical oxygen atoms in the $M$=V case will feel an
increasing flattening force, while for $M$=Cr, the force is going
to elongate the octahedron. Thus, in these two systems, the subtle
balance of competition between the electron-electron correlation
and the electron-lattice interaction may be a key factor to
determine their physical properties.

\begin{figure}
\centering
\includegraphics[width=0.8\textwidth]{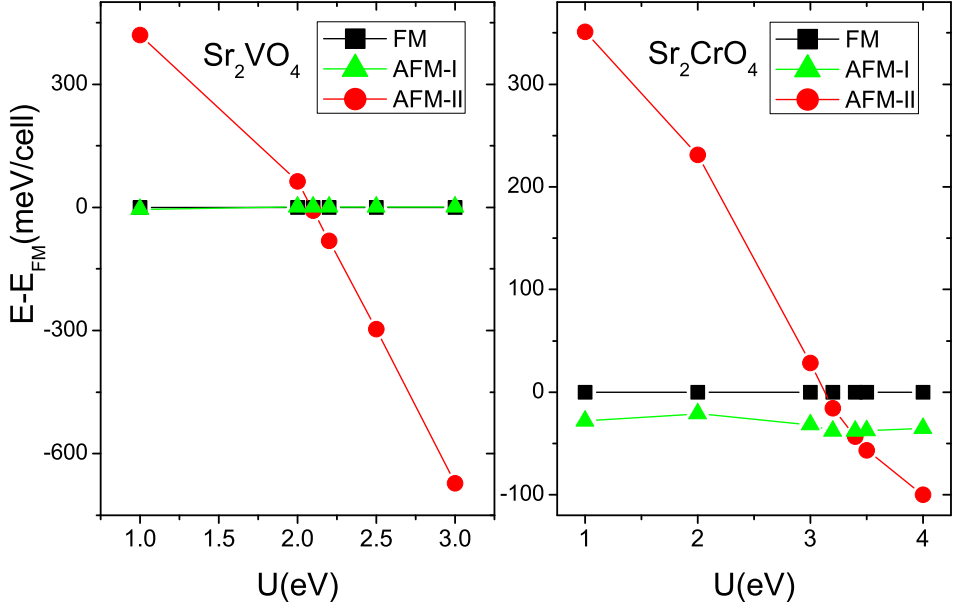}
\caption{(Color online) The variations of the total energies
relative to the FM state with the $U$ parameter for $M$=V (left
panel) and Cr (right panel).}\label{fig3}
\end{figure}

\begin{figure}
\centering
\includegraphics[width=0.8\textwidth]{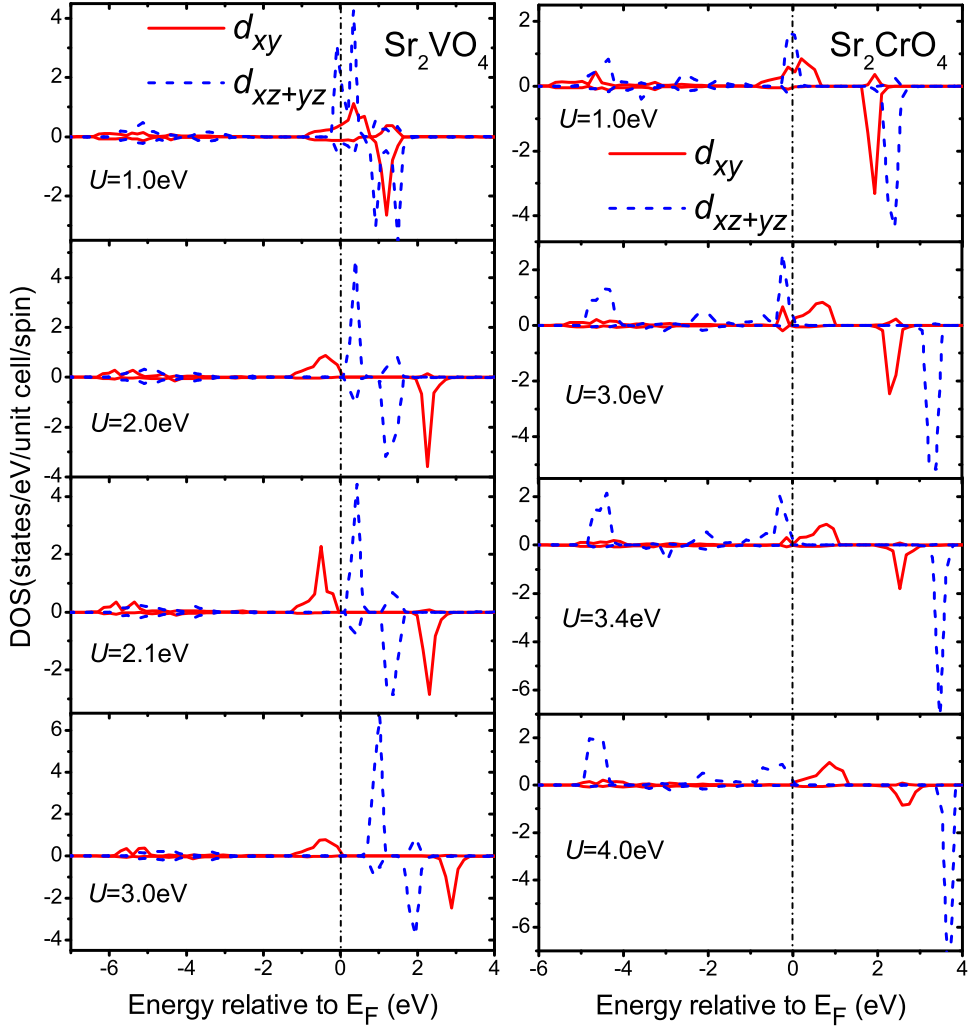}
\caption{(Color online) The variations of the $t_{2g}$ density of
states in the AFM-II state with the $U$ parameter for $M$=V (left
panel) and Cr (right panel). The vertical dash-dotted line
indicates the Fermi energy level.}\label{fig4}
\end{figure}

\begin{figure}
\centering
\includegraphics[width=0.8\textwidth]{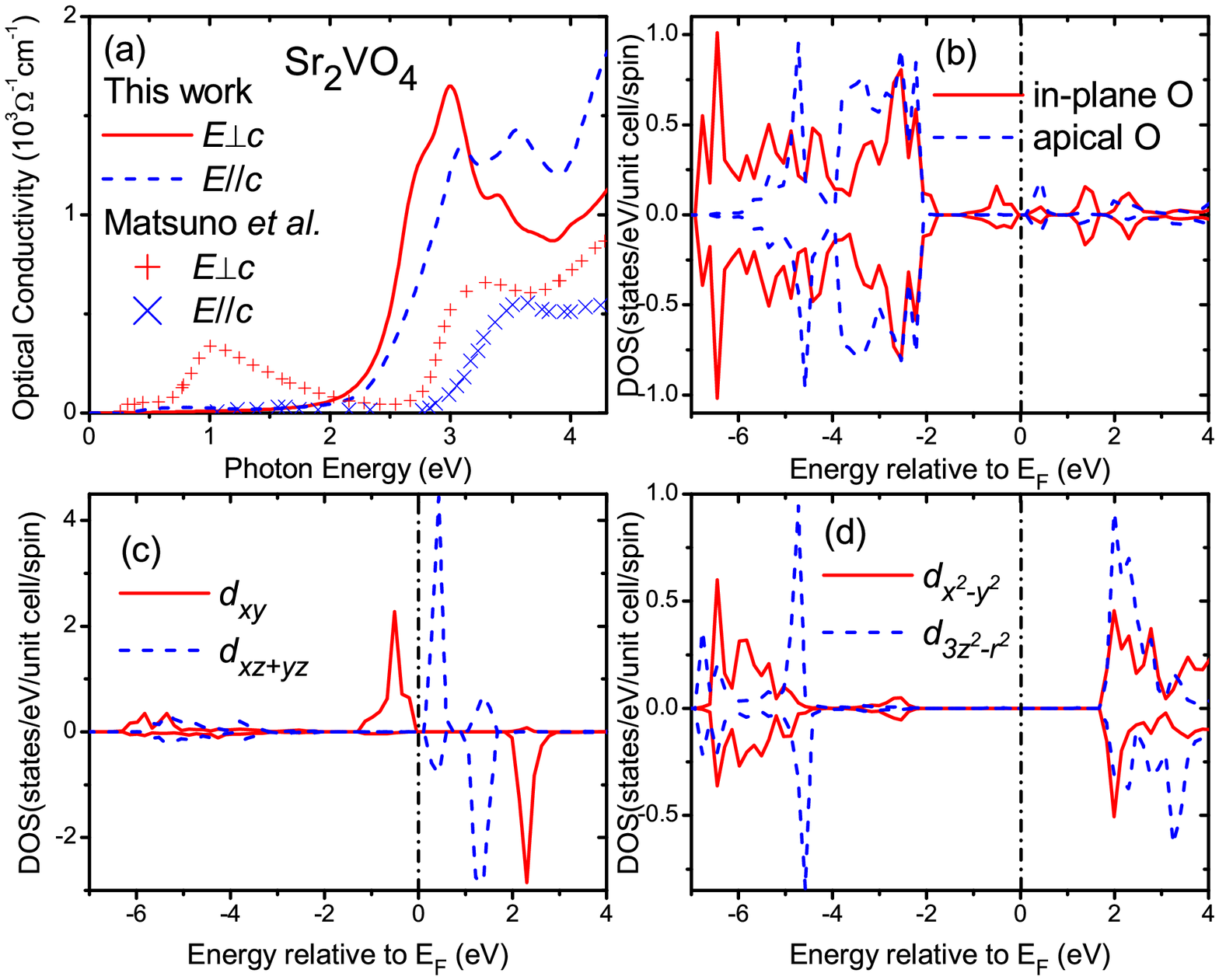}
\caption{(color online) (a) The experimental (symbols, Ref.
\onlinecite{srmo}) and calculated optical conductivity spectra
(lines) for Sr$_2$VO$_4$ in the AFM-II state with $U$=2.1 eV and
the corresponding partial DOS for (b) in-plane (solid line) and
apical (dashed line) oxygen $2p$ bands and (c) $t_{2g}$ and (d)
$e_{g}$ bands of V $3d$ electrons. The vertical dash-dotted line
indicates the Fermi energy level.}\label{fig5}
\end{figure}

\begin{figure}
\centering
\includegraphics[width=0.8\textwidth]{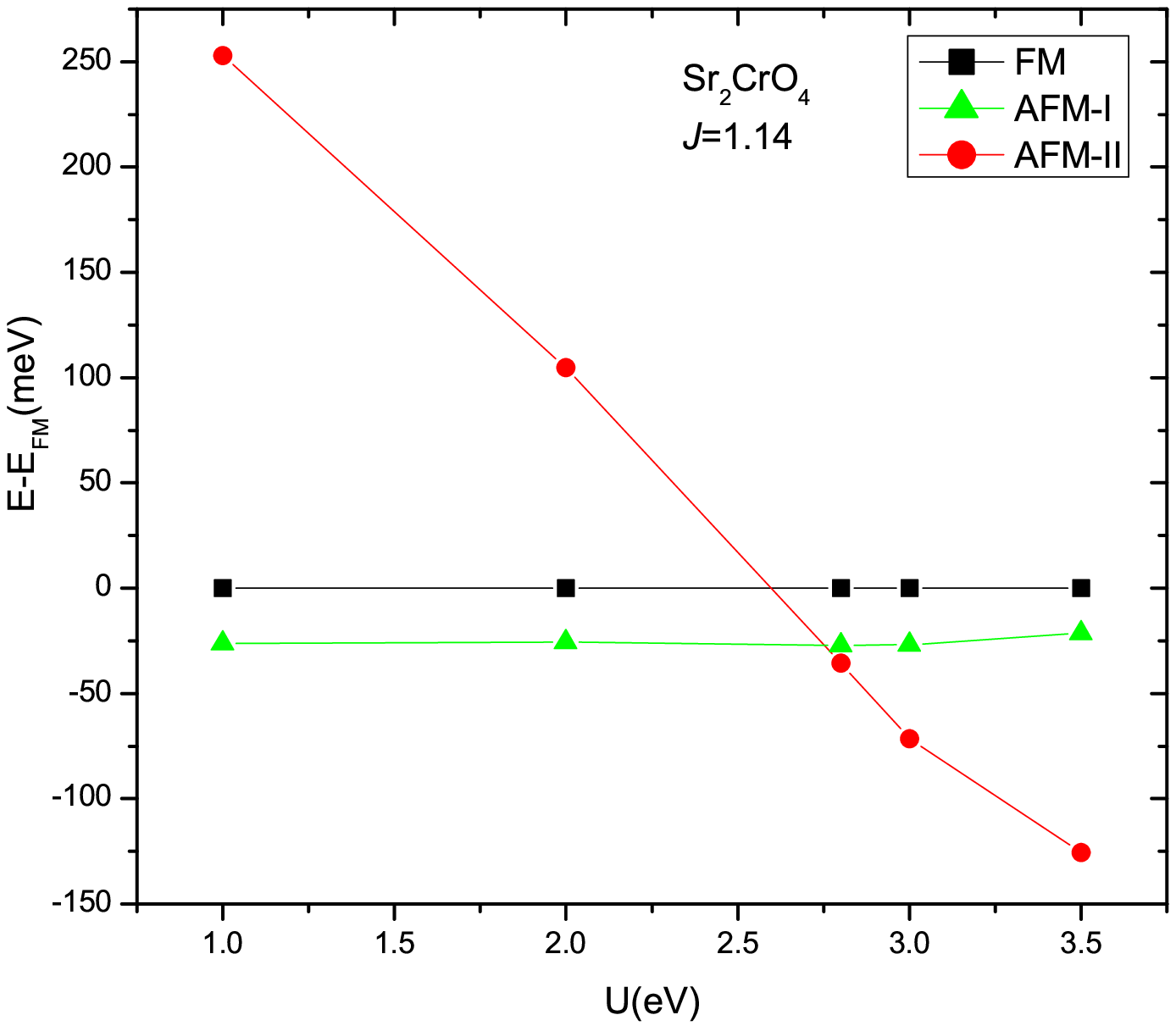}
\caption{(color online) The variation of the total energy relative
to the FM state with the $U$ parameter for $M$=Cr with the JT
distortion of $J$=1.14.}\label{fig6}
\end{figure}

In order to show the cooperative effect of the JT distortion and
$U$ in Sr$_2$CrO$_4$, we fixed $d_{ab}$ and artificially elongated
the oxygen octahedron to a degree with $J$=1.14. Again, the total
energies of various magnetic configurations relative to the FM one
with different $U$ values are shown in Fig. 6. Obviously, the
AFM-II state becomes the most stable when $U$ is larger than the
$U_c$ value of 2.74 eV, which is smaller than the original one of
3.36 eV. Due to the cooperation of JT distortion, the
metal-insulator transition is speeded up with the increasing of
$U$. The splitting of the triply degenerate $t_{2g}$ orbital is
key to such a transition and the realization of an AFM-II
insulator state for both Sr$_2$VO$_4$ and Sr$_2$CrO$_4$. Other
kind of lattice distortions, such as the rotation and/or tilting
of oxygen octahedra, which may cause the lattice deviation from
the tetragonal symmetry,\cite{braden} can also induce the energy
splitting in $t_{2g}$ orbitals and encourage orbital polarization
and hence be another driving force of the metal-insulator
transition. Here, it is noticed in the $M$=V case that the lattice
constants are in good agreement between the bulk and film samples
and the theoretical O$_a$ position is almost the same as the bulk
sample. The measured $a$ and $b$ lattice constants are nearly
equal, and the tetragonal symmetry is well kept. This means that
Sr$_2$VO$_4$ has no obvious orthorhombic distortion and a
contribution from the additional ligand field changes is unlikely.
In the $M$=Cr case, the orthorhombic strain, defined as
($b$-$a$)/($a$+$b$), is about 0.0112 in the samples, which is
quite large and comparable with that in Ca$_2$RuO$_4$, on average
being 0.012 25.\cite{braden} Then, for $M$=Cr, the lattice
distortions are expected to have a large influence on its magnetic
properties. But further investigation of these effects needs more
detailed geometric information, which is not available now
probably due to the difficulty in the synthesis of samples in
bulk. In the similar layered system Ca$_{2-x}$Sr$_x$RuO$_4$, Z.
Fang {\it et al.} has demonstrated that the detailed crystal
structure is crucial for interpretation of its magnetic phase
diagram and orbital physics.\cite{fang}

Similar to Sr$_2$VO$_4$, for $M$=Cr, the $U$ value of 3.4 eV is
taken for the optical conductivity calculation to ensure that the
AFM-II state is the lowest in energy and the energy difference
from the FM state is not too large, though at this $U$ value,
there is a small weight of DOS at the Fermi level. The theoretical
and experimental optical conductivity spectra are shown in Fig. 7.
Despite the metallic DOS feature, the obtained optical
conductivity spectra with both $E$$\perp$$c$ and $E$$\parallel$$c$
look like to have semiconducting features. One of the reasons for
this is that the contribution from the phenomenological Drude
component is not included in our calculation.\cite{srcoo} The
second one is the optical selection rule. According to this rule,
the optical transition from occupied $d_{xz+yz}$ to unoccupied
$d_{xy}$ is forbidden though the eigenvalue difference between
them is very small. In the case of $E$$\perp$$c$, intersite
$d-d^*$ transitions can occur and contribute to the peak near 3.5
eV, which could be explained by the $d_{xz+yz}$ DOS in Fig. 7(c).
The position of this peak is quite different from the experimental
analysis, where the peak around 1.0 eV is attributed to such a
Mott-Hubbard transition.\cite{srmo} The overestimation of this
peak originates from the $U$ value of 3.4 eV used in the
calculation, which locates the unoccupied upper Hubbard bands
$d_{xz+yz}$ as high as about 3.5 eV. The quite broad peak around
2.5 eV in the calculated $E$$\perp$$c$ spectrum comes from the so
called charge transfer transition between the in-plane oxygen $2p$
bands near -1.0 eV and the empty $d_{x^2-y^2}$ orbital at 1.5 eV.
However, in the corresponding experimental spectrum, such a peak
is around 2.0 eV. When $E$$\parallel$$c$, in the calculated
spectrum, the so-called charge transfer transition is around 3.5
eV, which comes from the transitions between apical oxygen $2p$
and Cr $d_{3z^2-r^2}$ orbitals. This peak position is also higher
than the corresponding experimental value by about 1.0 eV.

\begin{figure}
\centering
\includegraphics[width=0.8\textwidth]{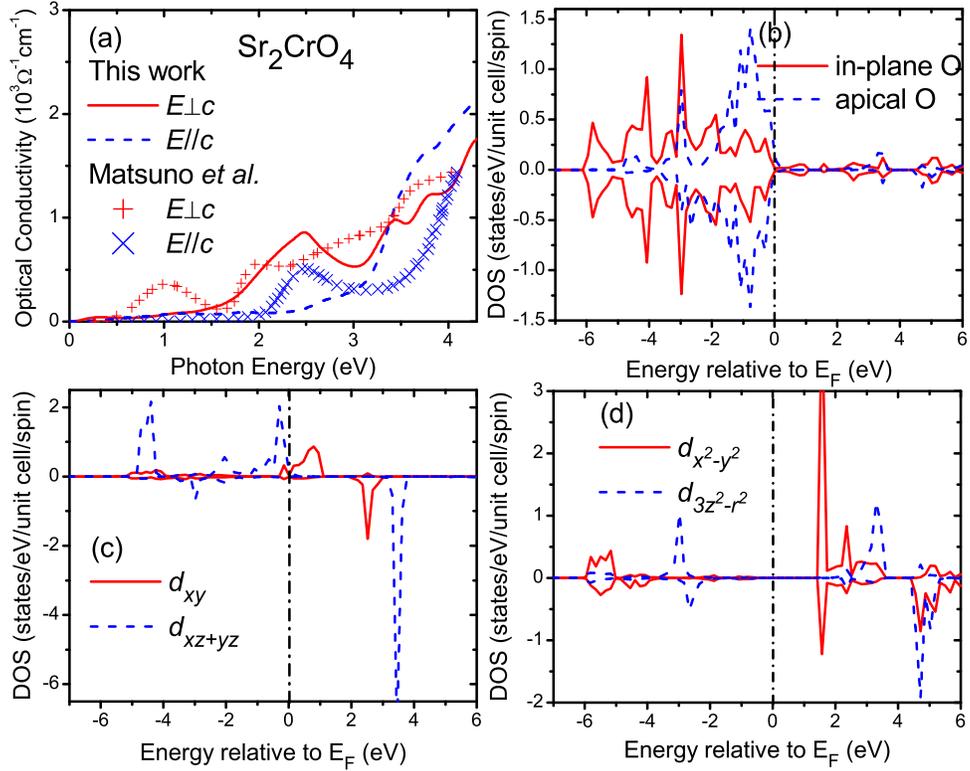}
\caption{(color online) (a) The experimental (symbols, Ref.
\onlinecite{srmo}) and calculated optical conductivity spectra
(lines) for Sr$_2$CrO$_4$ in the AFM-II state with $U$=3.4 eV and
the corresponding partial DOS for (b) in-plane (solid line) and
apical (dashed line) oxygen $2p$ bands and (c) $t_{2g}$ and (d)
$e_{g}$ bands of Cr $3d$ electrons. The vertical dash-dotted line
indicates the Fermi energy level.}\label{fig7}
\end{figure}

Therefore, compared with the experimental measurements,\cite{srmo}
the calculated optical conductivity spectra of Sr$_2$VO$_4$ and
Sr$_2$CrO$_4$ are not good enough due to the overestimation of the
$U$ parameter. As discussed above, just taking into account the
electron-electron correlation is not enough to give a good
description of the correct ground states of Sr$_2$VO$_4$ and
Sr$_2$CrO$_4$. Some other factors, such as the possible complex
spin and/or orbital ordering and the proper changes of
electron-lattice interaction, may be expected to make up the
deficiency, or even methods beyond the conventional LDA and
LDA+$U$ are needed. In the $M$=V $d^1$ case, we have discussed
that the contribution from the additional ligand field is
unlikely. With an approach combining the path-integral
renormalization group method with LDA calculations, Imai {\it et
al.} obtained a kind of AFM insulating state consistent with the
experiments and predicted a nontrivial orbital-stripe order in
Sr$_2$VO$_4$,\cite{path} while for $M$=Cr, the $d^2$ system, the
electron-lattice contribution is still possible. As shown in Fig.
6, the simple JT distortion is demonstrated to be quite efficient
although this may not be the real distortion in samples.

\subsection{Sr$_2$MnO$_4$}\label{sr2mno4}
The optimized structure parameter for Sr$_2$MnO$_4$ is 0.158 58
and 0.146 46 for O$_a$ and Sr. In the $M$=Mn case, both GGA and
GGA+$U$ calculations can predict the ground state to be the AFM-II
configuration, consistent with other studies.\cite{wang} Three $d$
electrons on the Mn$^{4+}$ ion occupy the three lower Hubbard
bands with the upper ones empty as shown by the DOS in Figs. 8(c)
and 8(d). In the calculated spectrum with $E$$\perp$$c$, the
optical gap comes from the transition between the occupied
in-plane oxygen $2p$ states and the empty Mn $d_{x^2-y^2}$ orbital
around 1.0 eV. Such a transition also contributes to the first
shoulder around 1.5 eV, and this shoulder corresponds to the first
peak in the experimental spectrum. The narrow peak around 2.2 eV
in the calculated $E$$\perp$$c$ spectrum can be interpreted from
panel (c) as the intersite $d-d^*$ transition, which is not
discussed in the experimental measurements.\cite{srmo} The peak
around 2.6 eV is due to the transition from the in-plane oxygen
$2p$ bands to the $t_{2g}$ and $d_{x^2-y^2}$ orbitals near 2.0 eV,
which corresponds to the second peak in the experimental spectrum.
When $E$$\parallel$$c$, the transition from the apical oxygen $2p$
bands to the Mn empty $d_{3z^2-r^2}$ orbital contributes most of
the peaks around 2.75 eV and 3.75 eV in the calculated spectrum.

\begin{figure}
\centering
\includegraphics[width=0.8\textwidth]{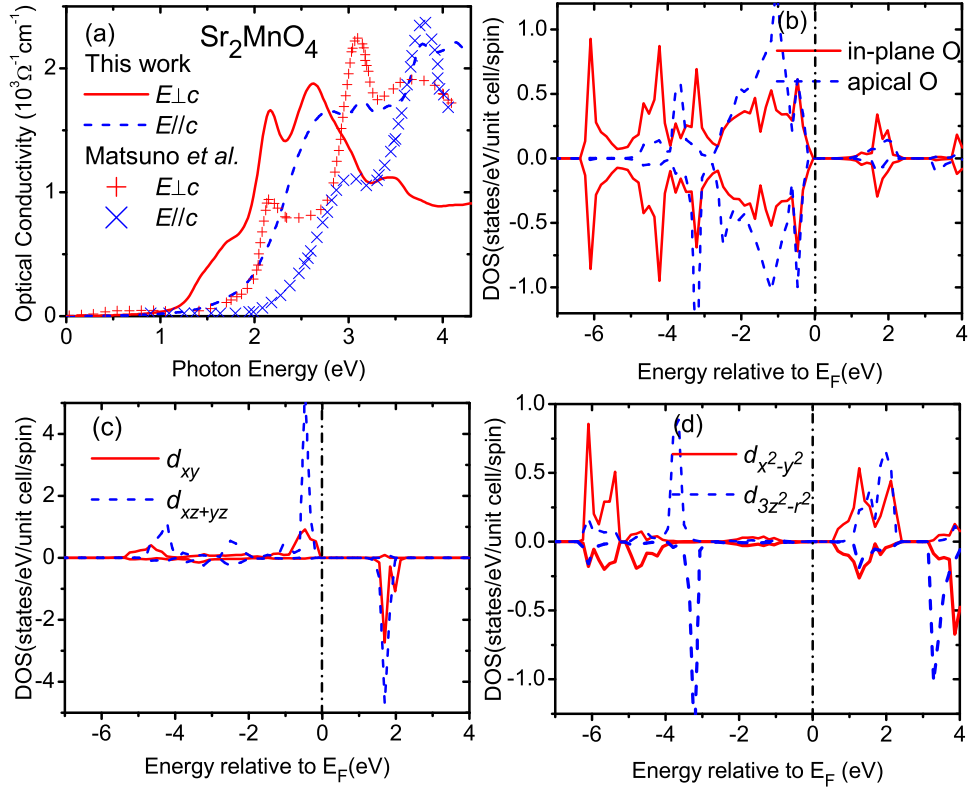}
\caption{(color online) (a) The experimental (symbols, from Ref.
\onlinecite{srmo}) and calculated (lines) optical conductivity
spectra for Sr$_2$MnO$_4$ in the AFM-II state within the GGA, and
the corresponding partial DOS for (b) in-plane (solid) and apical
(dashed) oxygen $2p$ bands and (c) $t_{2g}$ and (d) $e_{g}$ bands
of Mn $3d$ electrons. The vertical dash-dotted line indicates the
Fermi energy level.}\label{fig8}
\end{figure}

\section{Conclusion} \label{Conclusion}
For a series of layered perovskites Sr$_2$$M$O$_4$ ($M$=Ti, V, Cr,
and Mn), their electronic structures and the optical conductivity
spectra are studied by first principles calculations within the
GGA and GGA+$U$. The GGA calculation could successfully predict
the ground state for $M$=Ti and Mn, but failed for $M$=V and Cr.
In these two strongly correlated $t_{2g}$ systems, GGA+$U$
calculations show that the AFM-II state will have the lowest total
energy when $U$ is larger than a critical value $U_c$. But the
optical conductivity spectra indicate that in both cases, $U_c$ is
overestimated. Analysis indicates that some cooperative changes in
the ligand filed such as JT or some other lattice distortions
could reduce the $U_c$ value. These changes in the
electron-lattice interaction will induce further splitting of the
degenerate $t_{2g}$ orbital just as $U$ does, which is shown to be
critical for the metal-insulator transition and stabilization of
AFM-II configuration in these systems. In the $M$=V case, since
the geometric structure in bulk or thin films is well comparable
with the theoretical one and nearly has no orthorhombic
distortion, such a kind of additional contribution from lattice
distortion seems unlikely. In Sr$_2$CrO$_4$, it has quite a large
orthorhombic deviation from tetragonal symmetry in samples and a
simple JT distortion is demonstrated to be quite efficient in
decreasing the $U_c$ value although this may not be the real
distortion in samples. Further first-principles study of the
orthorhombic distortion effects needs much detailed experimental
geometric information, for which more experimental efforts are
expected.


\begin{acknowledgments}

The authors thank the staff of the Center for computational
Materials Science at the IMR for their support and the use of
Hitachi SR8000/64 supercomputing facilities. H.M.W. acknowledges
the critical reading of this work by J. Matsuno and valuable
discussion with Zhijian Wu and P. Murugan. X.G.W. acknowledges
support from the Natural Science Foundation of China under Grant
No. 10304007 and National Key Projects for Basic Researches of
China under Grant No. 2006CB0L1002.

\end{acknowledgments}



\end{document}